\documentclass[journal]{IEEEtran}
\ifCLASSINFOpdf
\else
   \usepackage[dvips]{graphicx}
\fi
\usepackage{url}
\usepackage{graphicx}
\usepackage{amsmath,bm}
\usepackage{amssymb}  
\usepackage{cite}        
 \usepackage{epstopdf} 
  \usepackage{xcolor} 
  \usepackage{subfigure} 
  \usepackage{amsthm}

\begin{document}

\title{Model-aided Deep Neural Network  for Source Number Detection}

\author{Yuwen Yang,  Feifei Gao, Cheng Qian, and Guisheng Liao
\thanks{Y. Yang and F. Gao are with  Institute for Artificial Intelligence Tsinghua University
(THUAI), State Key Lab of Intelligent Technologies and Systems, Beijing National Research Center for Information Science and
Technology (BNRist), Department of Automation, Tsinghua University, Beijing,
100084, P. R. China (email: yyw18@mails.tsinghua.edu.cn, feifeigao@ieee.org).}
\thanks{C. Qian is with the Department of Electrical and
Computer Engineering, University of Virginia, Charlottesville, VA 22904 USA
(e-mail: alextoqc@gmail.com). }
\thanks{G. Liao is with  the National Laboratory of Radar Signal Processing,
Collaborative Innovation Center of Information Sensing and Understanding,
Xidian University, Xi'an 710071, China (e-mail: liaogs@xidian.edu.cn).}
}

\markboth{xxx, April 2019}
{Shell \MakeLowercase{\textit{et al.}}: Bare Demo of IEEEtran.cls for IEEE Journals}
\maketitle

\begin{abstract}
Source number detection is a critical problem in array signal processing. Conventional model-driven methods e.g., Akaikes information criterion (AIC) and minimum description
length (MDL), suffer from severe performance degradation  when the number of snapshots is small or the signal-to-noise ratio (SNR) is low. In this paper, we exploit the model-aided based deep neural network (DNN) to  estimate the source number. Specifically, we first propose the eigenvalue based regression network (ERNet) and classification network (ECNet) to estimate the number of  non-coherent sources, where the eigenvalues of the received signal covariance matrix and the source number are used as the input and the supervise label of the networks, respectively.
Then, we extend the ERNet and ECNet for estimating the number of coherent sources, where the forward-backward spatial smoothing  (FBSS) scheme is adopted to improve
 the performance of ERNet and ECNet.
Numerical results demonstrate the outstanding performance of  ERNet and ECNet over the conventional AIC and MDL methods as well as their excellent generalization capability, which also shows their  great potentials for practical applications.
\end{abstract}


\IEEEpeerreviewmaketitle

\section{Introduction}
Estimation of  the source number from noisy
received data is a fundamental problem in array
signal processing and related applications, such as  radar \cite{khan2013autonomous}, acoustic signal processing  \cite{6819064}, wireless communications \cite{8354789}, Internet of Things (IoT) \cite{7305269}, etc.
In particular, most  direction-of-arrival (DOA) estimation algorithms (e.g., MUSIC and ESPRIT \cite{8212844}) rely on the prior knowledge o accurate source  number and would fail when such knowledge is unavailable.

In the last three decades, various methods have been proposed to estimate the number of sources, among which the  Akaikes information criterion (AIC) and the
minimum description length (MDL) are the two most popular
ones. In these two methods, the estimated number of sources is
 the value that minimizes the AIC or the MDL criterion \cite{williams1999detection}.
 However, AIC and MDL exhibit performance degradations
 when the number of snapshots is small or the signal-to-noise ratio (SNR) is low.
 Performance analyses and improvements of AIC and MDL
 are provided in \cite{5701798,4286946}.



In  practical systems, coherent sources would exist, e.g.,  multipath propagation and smart jammer scenarios.
Unfortunately, AIC and MDL related methods cannot be directly applied to the coherent source scenarios
due to the rank-deficient  covariance matrix.
One solution  \cite{article} is to employ forward-backward spatial smoothing (FBSS) to ``decorrelate'' the source signals.
By partitioning the whole array into sub-arrays and  averaging of the sub-array received covariance matrices, the equivalent source covariance matrix becomes nonsingular.
The FBSS scheme is  widely-used  since it can be regarded as a preprocessing before applying   AIC and MDL  \cite{article} as well as  DOA estimation algorithms.


Recently,  deep learning (DL) has drawn growing attentions for its great success
in various areas, such as computer vision,  automatic speech recognition and wireless communications  \cite{8663966,hemodelmodel,8672767,8680715,8795533}.
 Apart from the excellent learning capability, remarkable  generalization property and low computational complexity of DL are both attractive advantages that promote its applications.
In this paper, we exploit  the  deep neural network (DNN) with the aid of signal modeling information to  estimate the number of  sources. We first develop the eigenvalue based regression network (ERNet) and classification network (ECNet) for  non-coherent sources.
Then,  we propose  the FBSS based ERNet  and ECNet  to handle the case of coherent sources.
Simulation results show that the proposed networks
achieve significantly better performance than the conventional model-based methods (i.e., AIC, MDL, FBSS based AIC and FBSS based MDL), and
demonstrate the efficiency and remarkable generalization capability of the proposed networks.


\section{Signal Model}\label{secmodel}
Consider $K$    signals emitted from the far field impinging on an array of $M$ antennas, the $n$-th snapshot of the array can be written as
\begin{equation}
\bm r(n)=\bm A(\bm \theta) \bm s(n)+\bm w(n),
\end{equation}
where $(\cdot)^{T}$ denotes the transport  operator, $\bm s(n)=\left[s_{1}(n), s_{2}(n), \cdots, s_{K}(n)\right]^{\mathrm{T}}$ is the signal vector,
$\bm w(n)=\left[w_{1}(n), w_{2}(n), \cdots, w_{M}(n)\right]^{\mathrm{T}}$ is the noise vector,
$\bm \theta=\left[\theta_{1}, \theta_{2}, \cdots, \theta_{K}\right]^{\mathrm{T}}$ is the DOA vector of all $K$ users,
and  $\bm A(\theta)=\left[\bm a\left(\theta_{1}\right), \bm a\left(\theta_{2}\right), \ldots, \bm a\left(\theta_{K}\right)\right]$
is the $M \times K$ matrix of the steering vectors.
Suppose the number of snapshots  is $N$. The  covariance matrix of $\bm r(n)$  is estimated from the limited received data as
\begin{equation}
\hat{\bm R}=\frac{1}{N} \sum_{n=1}^{N} \bm r(n) {\bm r}^{H}(n).
\end{equation}

To estimate the number of  sources, we consider the following assumptions.
\begin{enumerate}
  \item The condition $\theta_{i}\ne \theta_{j}$  holds  for $\forall i\ne j $ with $i,j=1,2,\cdots,K$.
  \item The number of sensors is greater than the number of sources, i.e.,  $K < M$.
  \item 
      The noise vector $\bm w(k)$ is  assumed to be uncorrected with the source signals.
\end{enumerate}
Notice that we do not make any assumption about the statistical nature of
the signals  $\left\{ s_{n}(k)\right\}_{n=1}^{N}$ or the noise $\bm w(k)$.

\begin{figure}[!t]
\centering 
\includegraphics[width=80 mm]{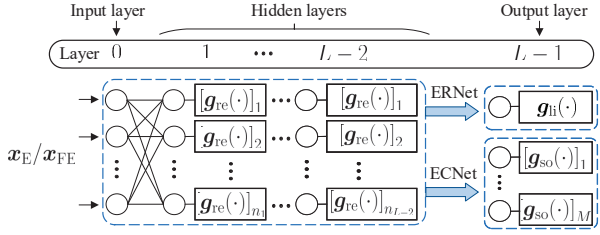}
\caption{ The architectures of ERNet and ECNet.}
\label{figdnn}       
\end{figure}
\section{Deep Learning based Source Number Detection}\label{secdll}
\subsection{Motivation of  Model-aided DNN }\label{secanaa}
 Based on information theory \cite{williams1999detection}, the   covariance matrix $\hat{\bm R}$
 contains the key information for the source number detection. One natural idea is to employ DNN to learn the source number directly from  the covariance matrix $\hat{\bm R}$. However, it is would not be an effective way to to so due to the information redundancy in  $\hat{\bm R}$ that contains all unknown information besides the number of sources (will be verified in simulations).

Although DL exhibits powerful learning capability in extracting features from massive data, model information, if can be well utilized, would provide significant enhancement in performance\cite{hemodelmodel}. To see this, let us express $\hat{\bm R}$  as
 \begin{equation}\label{equsd}
 \hat{\bm R}=\bm{U\Lambda}\bm U^{H}=\sum\limits_{m=1}^{M} \lambda_{m} \bm {u}_m  {\bm {u}_m}^H
 \end{equation}
where  $\lambda_{m}$ and $\bm {u}_m $ are the $m$-th eigenvalue  and corresponding eigenvector of  $\bm {\hat{R}}$, respectively. From the array signal processing theory \cite{8212844}, the eigenvectors would construct the signal subspace and the noise subspace, from which the ESPRIT and MUSIC algorithms can be applied to estimate the DOA, respectively. Nevertheless, it is clearly that eigenvectors themselves do not contain information of the number of the sources and hence would bring redundancy if being input to the DNN. Hence, instead of input the whole $ \hat{\bm R}$, we propose to only input the DNN the eigenvalues
 $\lambda_{1},\cdots,\lambda_{M}$, which is also consistent with the AIC and MDL approaches.
Notice that the eigenvalues $\lambda_{1},\cdots,\lambda_{M}$ are all real numbers, therefore,
 complex operations are not required in the proposed  ERNet and ECNet.

\subsection{Architectures of  ERNet and ECNet}
 The proposed ERNet and ECNet adopt  the fully-connected  neural network (FNN)  architecture
with ${L}$ layers, including  one input layer,  ${L}-2$ hidden layers, and one output layer as shown in Fig.~\ref{figdnn}. The input of both  ERNet and  ECNet is
 \begin{equation}\label{equndshn}
\bm{ x}_{\textrm{E}}= [\lambda_{1},\cdots,\lambda_{M}]^{T}.
\end{equation}
The output of  FNN is a cascade of nonlinear transformation of $ \bm{ x}_{\textrm{E}}$, i.e.,
\begin{equation}\label{equnn}
\bm{ \hat y}= \textrm{NET}\left(  \bm{ x}_{\textrm{E}}, \bm\Omega \right)=\bm f^{(L-1)}\left(\cdots \bm f^{(1)}\left(\bm{ x}_{\textrm{E}}\right)\right),
\end{equation}
where $L$ is the number of layers and $\bm\Omega\buildrel \Delta \over =\left\{\bm W^{(l)},\bm b^{(l)}\right\}_{l=1}^{L-1}$  is the network parameters to be trained.
Moreover, $\bm f^{(l)}$  is the nonlinear transformation function of the  $l$-th layer and can be written as
\begin{equation}\label{equnfd}
\bm f^{(l)}\left( \bm x \right) =\bm g^{(l)}\left(\bm W^{(l)}\bm x +\bm b^{(l)}\right),\  1\le l\le L-1,
\end{equation}
where $\bm W^{(l)}$  is the   weight matrix associated with the ($l-1$)-th and $l$-th layers, while $\bm b^{(l)}$  and $ \bm g^{(l)}$ are the bias vector and the activation function of the  $l$-th layer, respectively. The activation function for the hidden layers  is selected as
  the rectified linear unit (ReLU) function $\left[\bm g_{\textrm{re}}(\bm z)\right]_{p}=\max\{0,\left[\bm z\right]_{p}\}$, where $\left[\bm z\right]_p$ denotes the $p$th
entry of the vector $\bm z$, $\textrm{len}(\bm z)$ represents the length of   the vector $\bm z$, and $\ p= 1,2,\cdots,\textrm{len}(\bm z)$.

For ERNet, the supervise label is the number of sources, $K$; The  estimated number of sources, $\hat K_{\textrm{ER}}$, is the output of ERNet rounded to the nearest integer; The activation function
for the  output layer is  the linear function, i.e.,  $\bm g_{\textrm{li}} (\bm z) =\bm z$; The loss function  is the $L_2$ norm function, i.e.,
\begin{equation}\label{ewhadu}
   \textrm{Loss}_{L2}\left(\bm \Omega\right)=\frac{1}{{V}}\sum\limits_{v= 0}^{ V-1}
\left\|\hat{\bm{y}}(v)-{\bm{y}}(v)  \right\|^{2}_2 ,
\end{equation}
where $\left\| {\cdot} \right\|_2$  denotes the $L_2$  norm, $V$ is the batch size\footnote{Batch size is the number of samples in one training batch.}, and  $v$ denotes the index of the $v$-th  training sample.

For ECNet, the supervise label is the one-hot encoding vector\footnote{One-hot encoding vector related to $N$ and $M$ is an $M$-dimension vector with its $(N+1)$-th element being $1$ and other elements being $0$.} related to $K$ and $M$; The  estimated number of sources is the position index of the maximum element of the output vector $\hat {\bm y}_{\textrm{EC}}$, i.e., $\hat K_{\textrm{EC}}=\arg\max_{p}[\hat {\bm y}_{\textrm{EC}}]_{p}$-1; The activation function
for the  output layer is the Softmax function, i.e.,
\begin{equation}\label{equsd}
 \left[\bm g_{\textrm{so}}(\bm z)\right]_{p} =\frac{e^{\left[\bm z\right]_{p} }}{\sum_{p=1}^{p=\textrm{len}(\bm z)}e^{\left[\bm z\right]_{p} }}, \ p= 1,2,\cdots,\textrm{len}(\bm z);
\end{equation}
The loss function  is the categorical cross entropy function, i.e.,
\begin{equation}\label{sfgsfg}
   \textrm{Loss}_{\textrm{cc}}\left(\bm \Omega\right) = \frac{1}{{V}}\sum\limits_{v= 0}^{ V-1}
\sum\limits_{p= 1}^{ \textrm{len}(\bm y )} \left[{\bm{y}}(v)\right]_{p}\log \left[\hat{\bm{y}}(v)\right]_{p}.
\end{equation}

The DL based algorithm has two stages,  i.e., the off-line training and the on-line testing stages.
In the off-line training stage, the inputs and the supervise labels are collected as training samples.
The optimal $\bm \Omega$  can be obtained by minimizing the difference between the outputs and the labels through off-line training.
The adaptive moment estimation (ADAM)  algorithm  \cite{kingmaadam}  is adopted to minimize the
loss function until the loss converges, and then the optimal  $\bm \Omega$  can be obtained.
In the testing stage,  $\bm \Omega$ is fixed, and the network directly outputs the estimates of the labels based on the input data.

%
%

\begin{table*}
\centering
\caption{Comparison of computational complexity}
\begin{tabular}{|c|c|c|c|c|c|}
  \hline
  Methods & ERNet & ECNet&  AIC & MDL \\
    \hline
    Eigenvalue decomposition &Yes & Yes  & Yes &Yes \\
   The number of multiplications/divisions & $Mn_1+n_2$ & $M(n_1+n_2)$  & $M^2+7M$ & $M^2+7M$ \\
  The number of additions/subtractions & $n_1+n_2+1$ & $n_1+n_2+M$ & $0.5(M^2+M)$  & $0.5(M^2+M)$ \\
   The number of logarithmic operators & 0 & 0 &  $2M$ & $M$ \\
   The number of comparisons & 0 & $M$ &  $M$ & $M$ \\
  \hline
\end{tabular}
\label{tablestu}
\end{table*}

\subsection{Detection in the Case of Coherent Sources}
When the sources are coherent, which is the common in the real environment, the signal covariance matrix is rank-deficient, and therefore the performance of eigenvalue based methods degrades  \cite{7428571}. To solve the problem, the FBSS scheme is adopted \cite{article}, where
the the total array of $M$ antennas is divided forward/backward  into overlapping sub-arrays with size $M_0$.
The number of forward/backward  sub-arrays is $T=M-M_0+1$.
Note that the $M_0> K$ and $T\ge K$ should be satisfied as proved in \cite{7428571}.
Then, the forward/backward averaged covariance matrix is given by \cite{article}
\begin{equation}
\hat{\bm R}_{\textrm{ave}}=\frac{1}{2T} \sum_{t=1}^{T} \left(\hat{{\bm R}}_{\textrm{f}}^{(t)}+ \hat{{\bm R}}_{\textrm{b}}^{(t)}\right),
\end{equation}
where   $\hat{{\bm R}}_{\textrm{f}}^{(t)}$ and $ \hat{{\bm R}}_{\textrm{b}}^{(t)}$ are the
covariance matrices of the received signals that are estimated from the $t$-th ($1\le t\le T$) forward/backward sub-arrays, respectively. Details about FBSS scheme can be found in \cite{7428571}.

With the FBSS scheme, we refine the input of  ERNet and  ECNet as
 \begin{equation}\label{equndshn}
\bm{ x}_{\textrm{FE}}= [\lambda_{\textrm{ave},1},\cdots,\lambda_{\textrm{ave},M_0}]^{T},
\end{equation}
where $\lambda_{\textrm{ave},1},\cdots,\lambda_{\textrm{ave},M_0}$ are the eigenvalues of the $\hat{\bm R}_{\textrm{ave}}$, and the training approach
is the same.


\subsection{Complexity Analysis}
Denote $n_l$ as the number of  neurons in the $l$-th layer, and the number of layers, $L$, is set to be 4 in the proposed networks. The computational costs of  ERNet, ECNet,  AIC and MDL methods are compared, as listed  in  Tab.~\ref{tablestu}.
Since the eigenvalue decomposition is required in all the above-mentioned method, the  computational cost of the eigenvalue decomposition  is omitted in the  Tab.~\ref{tablestu}.
 Note that the  Softmax in ECNet is unnecessary in the testing  stage since the estimate of $K$  can be directly obtained by finding the index of the maximum element in the output vector.
 Therefore, the computational cost of Softmax is not included in ECNet.
 As shown in the Tab.~\ref{tablestu}, the  complexity of both AIC and MDL estimators is $O(M^2)$ while the the  complexity of ERNet and  ECNet is $O(M)$, which implies both ERNet and ECNet have lower complexities than AIC/MDL at the cost of off-line training.


\begin{figure}[!t]
\centering 
\includegraphics[width=88 mm]{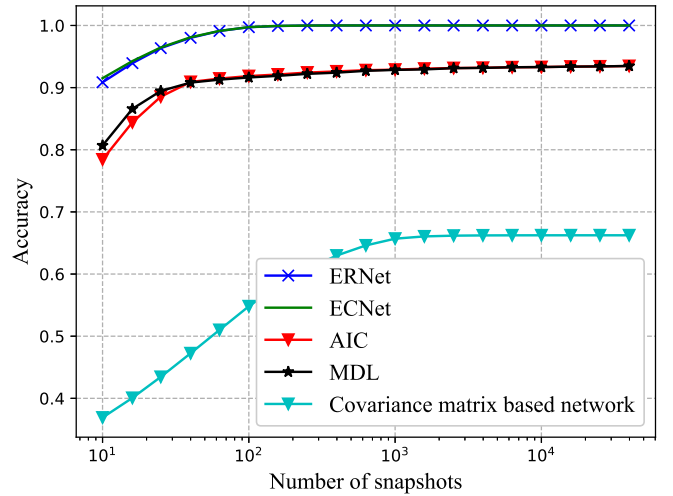}
\caption{The accuracy performance versus the number of snapshots, non-coherent sources, SNR=5 dB.}
\label{figsnap}       
\end{figure}
\section{Simulation Results}\label{secsim}
In this section, a uniform linear array (ULA) with 10 antennas is used to received the  signals.
Unless otherwise specified, the simulation parameters are set as follows:
The number of snapshots $N$ is 20;
The number of sources $K$ is generated randomly over $[0,5 ]$;
The direction of each source is generated randomly over $[0,2\pi]$; The number of neurons in the hidden layer are all (8,8) for ERNet and ECNet.
Keras 2.2.0 is employed as the DL framework.
The initial learning rate of the ADAM algorithm is 0.001. The batch size is 128.
The  parameters of all the networks are initialized as truncated normal variables\footnote{The truncated normal distribution is a normal distribution bounded by two standard deviations from the mean.} with normalized variance\footnote{The  weights of  neurons in the $l$-th layer are initialized as truncated normal   variables with  variance   $1/n_l$.}.
The number of training samples is 8,000, and the number of epochs is 400.

It needs to be mentioned that the ERNet and ECNet  are both trained at varying SNRs rather than trained at each SNR separately. This is because that training and testing the networks at each  SNR separately is unpractical.
 In real systems, SNR tends to be varying or even unknown, and therefore it is almost impossible to obtain perfect training and testing sets, i.e.,
 the  statistics mismatches between the testing and training sets is unavoidable.
 By exploiting the generalization capability of DNN,
 we  train the networks  with samples generated at different SNRs and then test  the networks with samples generated at a fixed SNR.
 In this way, we can directly collect samples generated with different SNRs as training samples without the  knowledge of SNRs and the networks only need one-time training, which is more feasible than training at each SNR in practical applications.
 To generate a single training sample, the value of SNR is set as a number randomly selected from $[0,40]$dB.

\subsection{Source Number of Non-coherent Sources}

In the case of non-coherent sources, AIC and MDL estimators \cite{williams1999detection}
are used as benchmarks.

Fig.~\ref{figsnap} depicts the accuracy performance of the ERNet, ECNet,  AIC and MDL versus the number of snapshots, where SNR is set to be 5 dB in the testing stage.
As shown in  Fig. \ref{figsnap}, AIC and MDL exhibit similar performance while the  ERNet and ECNet both
significantly outperform the AIC and MDL estimators. Besides, when the number of snapshots  is less than 20, the  ECNet achieves slightly better performance than ERNet. When the number of snapshots goes beyond 100, both ERNet and ECNet almost yield perfect number detection, even when SNR is as low as 5 dB.
Besides, the covariance matrix based network directly estimates the  source number based on $\hat{\bm R}$ and performs much worse than MDL and AIC, which validates the analyses in Section~\ref{secanaa}.


\begin{figure}[!t]
\centering 
\includegraphics[width=88 mm]{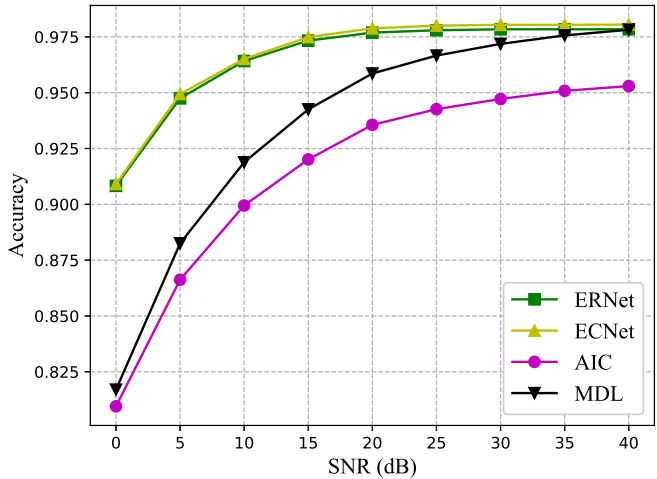}
\caption{The accuracy performance versus SNR, non-coherent sources,  $ K$=100.}
\label{figsnr}       
\end{figure}

Fig.~\ref{figsnr} compares the  accuracy performance of the ERNet, ECNet, AIC and MDL versus SNR when the number snapshots is set as 20.
 As shown in  Fig.~\ref{figsnr},  the performance of ERNet and ECNet is consistently good in the low  SNR region while ECNet achieves better performance than ERNet when SNR is higher than 20 dB.
Both ERNet and ECNet significantly outperform the AIC and MDL estimators, especially when SNR is lower than 20 dB, which demonstrates the remarkable superiority
 and the excellent generalization capability of  ERNet/ECNet.

\subsection{Source Number of Coherent Sources}
In the case of coherent sources, the FBSS based  AIC, and  the FBSS based MDL methods \cite{article}
are used as benchmarks. To generate a single training sample,
the number of coherent sources is generated randomly over $[0,K\!-\!1]$, and each coherent source is set to be identical with one of independent sources (randomly selected).
 The  size  of  sub-arrays is 5.
It should be noted that when the number of coherent sources is $K-1$, the sources
are non-coherent. 

\begin{figure}[!t]
\centering 
\includegraphics[width=88 mm]{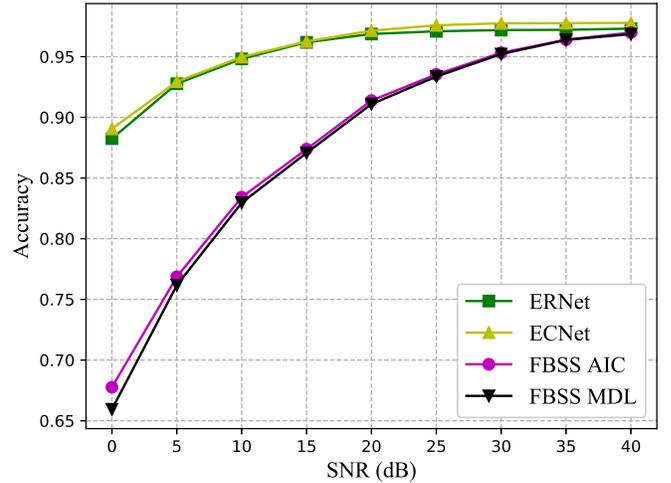}
\caption{The accuracy performance versus SNR, coherent sources, $ K$=100.}
\label{figcoherent}       
\end{figure}

As shown in Fig.~\ref{figcoherent}, the FBSS based AIC achieves slightly better performance than the FBSS based MDL when SNR is lower than 5 dB while  both ERNet and ECNet significantly outperform the FBSS based  AIC and MDL, especially when SNR is lower than 25 dB.
When SNR is 0 dB, the accuracy of the FBSS based AIC and MDL is about 0.7 while the accuracy of both ERNet and ECNet is higher than 0.95,
which shows the effectiveness of ERNet/ECNet in the case of coherent sources.
Furthermore,  ECNet achieves better performance than ERNet, especially when SNR is lower than 5dB or higher than 20 dB.

Moreover,  it can be observed from Figs.~\ref{figsnr} and \ref{figcoherent} that  when SNR is higher than 20 dB, the proposed networks achieve similar performance in both the non-coherent and  coherent cases.
While in  the low SNR region,  the proposed networks  achieve better performance in the non-coherent case than in the coherent case.

%

\section{Conclusion}
In this paper, we proposed the model-aided data-driven networks, i.e., ERNet and ECNet, to estimate both the number of  non-coherent sources and coherent sources.
The comparisons of computational complexities among the proposed and the conventional model-driven methods were presented in closed-form.
Simulation results have demonstrated that the proposed ERNet and  ECNet significantly outperform the conventional model driven methods  and have remarkable  generalization capability with respect to SNRs. Both the above properties make ERNet and ECNet great candidates  in real-world applications.

\bibliographystyle{IEEEbib}
\bibliography{References}

\begin{thebibliography}{10}

\bibitem{khan2013autonomous}
M.~Khan, K.~Iftekharuddin, E.~McCracken, K.~Islam, S.~Bhurtel, L.~Wang, and
  R.~Kozma,
\newblock ``Autonomous wireless radar sensor mote for target material
  classification,''
\newblock {\em Digital Signal Process.}, vol. 23, no. 3, pp. 722--735, 2013.

\bibitem{6819064}
Y.~{Wu}, Z.~{Hu}, H.~{Luo}, and Y.~{Hu},
\newblock ``Source number detectability by an acoustic vector sensor linear
  array and performance analysis,''
\newblock {\em IEEE J. Oceanic Engineering}, vol. 39, no. 4, pp. 769--778, Oct.
  2014.

\bibitem{8354789}
B.~{Wang}, F.~{Gao}, S.~{Jin}, H.~{Lin}, and G.~Y. {Li},
\newblock ``Spatial- and frequency-wideband effects in millimeter-wave massive
  {MIMO} systems,''
\newblock {\em IEEE Trans. Signal Process.}, vol. 66, no. 13, pp. 3393--3406,
  Jul. 2018.

\bibitem{7305269}
Y.~{Mohamedatni}, B.~{Fergani}, J.~. {Laheurte}, and B.~{Poussot},
\newblock ``{DOA} estimation techniques applied to {RFID} tags using receiving
  uniform linear array,''
\newblock in {\em IEEE Int. Symp. Antennas and Propagation USNC/URSI National
  Radio Science Meeting}, Vancouver, Canada, Jul. 2015, pp. 1760--1761.

\bibitem{8212844}
B.~{Vikas} and D.~{Vakula},
\newblock ``Performance comparision of {MUSIC} and {ESPRIT} algorithms in
  presence of coherent signals for {DoA} estimation,''
\newblock in {\em Proc. Int. Conf. Electron., Commun. Aerospace Technol.
  (ICECA)}, Apr. 2017, vol.~2, pp. 403--405.

\bibitem{williams1999detection}
D.~Williams and V.~Madisetti,
\newblock ``Detection: Determining the number of sources,''
\newblock {\em Digital Signal Processing Handbook}, 1999.

\bibitem{5701798}
Q.~{Ding} and S.~{Kay},
\newblock ``Inconsistency of the {MDL}: On the performance of model order
  selection criteria with increasing signal-to-noise ratio,''
\newblock {\em IEEE Trans. Signal Process.}, vol. 59, no. 5, pp. 1959--1969,
  May 2011.

\bibitem{4286946}
L.~{Huang} and S.~{Wu},
\newblock ``Low-complexity {MDL} method for accurate source enumeration,''
\newblock {\em IEEE Signal Process. Lett.}, vol. 14, no. 9, pp. 581--584, Sep.
  2007.

\bibitem{article}
S.~Shirvani Moghaddam and S.~Jalaei,
\newblock ``Determining the number of coherent sources using {FBSS}-based
  methods,''
\newblock {\em Frontiers in Science}, vol. 2, pp. 203--208, Dec. 2012.

\bibitem{8663966}
Z.~{Qin}, H.~{Ye}, G.~Y. {Li}, and B.~F. {Juang},
\newblock ``Deep learning in physical layer communications,''
\newblock {\em IEEE Wireless Commun.}, vol. 26, no. 2, pp. 93--99, Apr. 2019.

\bibitem{hemodelmodel}
H.~{He}, S.~{Jin}, C.~{Wen}, F.~{Gao}, G.~{Ye Li}, and Z.~{Xu},
\newblock ``Model-driven deep learning for physical layer communications,''
\newblock {\em IEEE Wireless Commun. (early access)}, pp. 1--7, 2019.

\bibitem{8672767}
Y.~{Yang}, F.~{Gao}, X.~{Ma}, and S.~{Zhang},
\newblock ``Deep learning-based channel estimation for doubly selective fading
  channels,''
\newblock {\em IEEE Access}, vol. 7, pp. 36579--36589, Mar. 2019.

\bibitem{8680715}
Z.~{Jia}, W.~{Cheng}, and H.~{Zhang},
\newblock ``A partial learning based detection scheme for massive {MIMO},''
\newblock {\em IEEE Wireless Commun. Lett.}, pp. 1--1, Apr. 2019.

\bibitem{8795533}
Y.~{Yang}, F.~{Gao}, G.~Y. {Li}, and M.~{Jian},
\newblock ``Deep learning based downlink channel prediction for {FDD} massive
  {MIMO} system,''
\newblock {\em IEEE Commun. Lett.}, pp. 1--1, 2019.

\bibitem{kingmaadam}
D.~Kingma and J.~Ba,
\newblock ``Adam: A method for stochastic optimization,''
\newblock {\em arXiv preprint arXiv:1412.6980}, 2014.

\bibitem{7428571}
S.~{Li} and B.~{Lin},
\newblock ``On spatial smoothing for direction-of-arrival estimation of
  coherent signals in impulsive noise,''
\newblock in {\em Proc. IEEE Adv. Inf. Technol., Electron. Automation Control
  Conf. (IAEAC)}, Chongqing, China, 2015, pp. 339--343.

\end{thebibliography}
\end{document}